\documentclass{article}
\usepackage{spconf,amsmath,graphicx}
\usepackage{tchdr}
\usepackage{layouts}

\usepackage[style=ieee,
			backend=bibtex,
			sorting=nyt]{biblatex}
\def\x{{\mathbf x}}
\def\L{{\cal L}}
\newcommand{\DI}{\text{DI}}
\newcommand{\ADI}{\text{ADI}}
\newcommand{\I}{\text{I}} 

\title{Time-Varying Interaction Estimation using Ensemble Methods}
\name{Brandon Oselio, Amir Sadeghian, Silvio Savarese, Alfred Hero\thanks{We
    acknowledge the support of USAF grant FA8650-15-D-1845 and US Army Research Office grant W911NF-15-1-0479.}}
\address{University of Michigan \\
  EECS Department \\
  1301 Beal Ave, Ann Arbor, MI 48109}
\address{University of Michigan \\
  EECS Departmest \\
  1301 Beal Ave, Ann Arbor, MI 48109}

\twoauthors
  {Brandon Oselio, Alfred Hero\sthanks{We
    acknowledge the support of USAF grant FA8650-15-D-1845 and US Army Research Office grant W911NF-15-1-0479}}%
	{University of Michigan \\
  EECS Department \\
  1301 Beal Ave, Ann Arbor, MI 48109}
  {Amir Sadeghian, Silvio Savarese}%
	{Stanford University\\
	Computer Science Department\\
	353 Serra Mall, Stanford, CA 94305}
\begin{document}
\maketitle
\begin{abstract}
Directed information (DI) is a useful tool to explore time-directed interactions in multivariate
data. However, as originally formulated
DI is not well suited to interactions that change over time. In previous work,
adaptive directed information was introduced to accommodate non-stationarity,
while still preserving the utility of DI to discover complex dependencies
between entities. There are many design decisions and parameters that are crucial to
the effectiveness of ADI. Here, we apply ideas from ensemble learning in order
to alleviate this issue, allowing for a more robust estimator
for exploratory data analysis. We apply these techniques to interaction
estimation in a crowded scene, utilizing the Stanford drone dataset as an example.
\end{abstract}
\begin{keywords}
directed information, adaptive directed information, temporal modeling, data
exploration, interaction mining
\end{keywords}

\section{INTRODUCTION}
\label{sec:intro}
The study of interactions among entities of interest encompasses a broad
array of applications and is crucial to understanding complex processes. Often
times, we are interested in the directionality over time of these
relationships. Examples include social influence estimation~\cite{OselioLH18,
  OselioH17, quinn2015directed}, entity interaction in
video~\cite{chen2014shrinkage},  and biological recording
analysis, such as EEG~\cite{chen2014eeg, quinn2011estimating}. These interactions
can also be used to summarize highly complex data topology, allow
analysts to obtain a qualitative snapshot of the temporal interactions of the
data, and make better informed decisions based on these simplified representations.

One tool that allows for the extraction of interactions is called directed
information (DI). Originally created to analyze an information-theoretic channel
with feedback, DI has been used in many contexts to estimate
directed relationships between entities, including genetic data and social data.
One deficiency of directed information is its
inflexibility with respect to time-varying distributions~\cite{OselioH17, OselioH16}. Adaptive directed
information (ADI) was developed as an extension of directed information to better
track changes in relationships over time.

In this paper, we address some of the issues associated with using ADI.
Specifically, ADI requires a choice of filter and corresponding filter
parameters, and the quality of the resulting interaction estimate is not
generally robust to these choices. In addition, simple filters may have
difficulty adapting to both abrupt changes in interaction, as well as
slowly time-varying systems. An estimate that is able to
accomplish both smoothing over time, as well as the ability to adapt to abrupt
changes in interactivity quickly is desired.

In this paper, a form of ensemble
learning is used to improve interaction estimation with ADI. Specifically, following~\cite{herbster1998tracking,
  shalizi2011adapting}, we generate a filter that is a convex combination of simpler
filters with different parameter specifications and whose weights are dependent
on the data. In order to address the possibility of abrupt changes in the
system, a growing ensemble of estimators is used
to account for these changes in interactivity.

The proposed ADI estimator is applied to interaction estimation in a
crowded scene, utilizing video from the Stanford drone dataset~\cite{robicquet2016learning}. Utilizing a dynamic covariance model, the ADI is estimated and
used to uncover interesting phenomena in specific scenes across the Stanford
campus.

The paper is organized as follows: Sec.~\ref{sec:rel_work} discusses related
work.
Sec.~\ref{sec:theory} introduces the
mathematical concepts of DI and ADI, and introduces our ensemble estimator.
Sec.~\ref{sec:agent_model} introduces the dynamic
covariance model used to estimate ADI. Sec.~\ref{sec:exp} discusses the results on
the Stanford Video Dataset. Finally, Sec.~\ref{sec:concl} concludes the paper.

\section{Related Work}
\label{sec:rel_work}
Directed information has been studied in the context of theory and applications. Estimators for DI have been proposed for the case of a finite or countably
infinite feature space~\cite{quinn2011estimating, jiao2013universal,
  liu2009directed}. Most, if not all, estimators use the stationary Markov
assumption, including plugin estimators~\cite{OselioH16, OselioH17}. Directed
information has been used in many contexts, including EEG
analysis~\cite{chen2014eeg}, neural spike trains~\cite{quinn2011estimating}, and
social influence analysis~\cite{OselioH16, OselioH17}.
Changepoint detection methods~\cite{aminikhanghahi2017survey} is one approach to
track time-varying data, and parametric as well as non-parametric methods
exist. However, with few exceptions, e.g., ~\cite{banerjee2018quickest} these
methods are mostly univariate and often require a parametric model or use simple
moment-based statistics that do not capture dependency.

Other methods of influence estimation have been studied, particularly in the
context of i.i.d. observations; examples include
glasso~\cite{friedman2008sparse} and hub discovery-type
methods~\cite{hero2012hub}. In addition, semi-parametric extensions of these
models have been created for non-Gaussian data~\cite{liu2009nonparanormal}.
The family of directed information measures and in
particular ADI is concerned with directionality in time and with more
complicated time-varying signals. In this paper, we assume a parametric
multivariate Gaussian model, which is appropriate for the particular dataset.

The ensemble method used stems from the prediction with multiple experts, a
popular problem in machine learning~\cite{cesa2006prediction,
  shalizi2011adapting, herbster1998tracking}. Here, we use these techniques for
smoothing.

\section{ADI and Ensemble Estimation}
\label{sec:theory}
\subsection{Definition of DI and ADI}
\label{ssec:defs}
We begin with some notation. We assume  that we have
$1, 2, \ldots, N$ entities each with features $\bX^i_{1:T} = \left[  X^i_1, X^i_2,
  \ldots, X^i_T \right]$. In
this paper, $X^i_t \in \reals^d$. Directed information between $X^i$ and $X^j$ is defined
as follows:
\begin{equation}
  \label{eq:di}
\DI(\bX_{1:T}^i \rightarrow \bX_{1:T}^j) = \sum_{t=1}^T \I(\bX_{1:t}^i;~X^j_{t} | \bX^j_{1:t-1}),
\end{equation}
where $\I(X;Y|Z)$ is the Shannon conditional mutual information. Many interesting
conservation properties have been derived for directed information, including a
close connection to the standard Shannon mutual information; these will not be
repeated here, but the reader is referred to
papers~\cite{massey1990causality, massey2005conservation, amblard2011directed}. When
considering the asymptotic behavior of $\DI$ for stationary processes, one
defines the directed information rate:
$$
\overline{\DI}(\bX^i \rightarrow \bX^j) = \lim_{T \rightarrow \infty} \frac{1}{T}\DI(\bX_{1:T}^i \rightarrow \bX_{1:T}^j).
$$
If we assume that the entities form a $k$-Markov process, then $\I(\bX_{1:t}^i;X^j_{t}
| \bX^j_{1:t-1}) = \I(\bX_{1:t}^i;X^j_{t}| \bX^j_{t-k:t-1})$.

When stationarity cannot be assumed, then the traditional definition of
$\overline{\DI}$ is inapplicable. However, the instantaneous DI summand
of~(\ref{eq:di}) retains valuable
information about temporal interactivity of the entities $i$ and $j$.
In~\cite{OselioH17}, we proposed to adaptively estimate this quantity using adaptive directed information
(ADI), which is defined as follows:
$$
\ADI(\bX_{1:T}^i \rightarrow \bX_{1:T}^j) = \sum_{t=1}^T g(t,T)\I(\bX_{1:t}^i;X^j_{t} | \bX^j_{1:t-1}),
$$
where $g(t,T)$ is a user-defined taper function. In past work~\cite{OselioH17}, the focus has been on
the exponential filter $g(t,T) = \alpha(1 - \alpha)^{t-T},$ so that ADI obeys
the recursive update:
$$
\ADI_{1:t}^{i \rightarrow j} = \alpha \I(\bX^i;~X^j_{t} | \bX^j_{1:t-1}) + (1 -
\alpha) \ADI_{1:t-1}^{i \rightarrow j},
$$
where $\ADI_{1:t}^{i \rightarrow j} = \ADI(\bX_{1:t}^i \rightarrow \bX_{1:t}^j)$.
However, the parameter $\alpha$ of the exponential filter must be tuned according to the
specific application. The goal of the this paper is to improve the
robustness of ADI when the underlying state is unknown and rapidly changing. In order to
accomplish this, an ensemble filter is defined:
\begin{equation}
  \label{eq:adi_ensemble}
g^*(t,T) = \frac{\sum_{i=1}^{n_t} w_{i,t} g_i(t,T;t_0)}{\sum_{i=1}^{n_t}w_i},
\end{equation}
where $g_i(t,T)$ are ``base filters'' with different parameter specifications. Implicitly,
the weights $w_i$ are allowed to depend on past data. Further, the number of
base filters included in the ensemble ($n_t$) is allowed to grow with $t$, and filter
functions will be causal, i.e., $g(t,T; t_0) = 0$ for $t < t_0$.

\subsection{Expanding Fixed Shares of Estimation}
\label{ssec:ensemble}

We apply an ensemble method based on the simple fixed shares
algorithm~\cite{herbster1998tracking}, which was
originally introduced in~\cite{shalizi2011adapting}. 

A set of base filter functions is defined, $G = \{g_1, \ldots,
g_k\}$ along with a parameter $\tau$ which
defines the rate at which new filters are introduced into~\ref{eq:adi_ensemble}

At each time $t$, an estimate $\hat{\I}(\bX_{1:t}^i;~X^j_{t} |
\bX^j_{1:t-1})$ is obtained and used to both update the weights $w_i$ and to
update the ADI estimate.

The weights $w_i$ are updated in a similar manner to~\cite{shalizi2011adapting}:
$$
v_{i,t} = w_{i,t-1}e^{-\gamma(y_{i,t} - i_t)^2}, w_{i,t} = (1 - \beta)v_{i,t} + \frac{\beta}{n_t}\sum_{i=1}^{n_t}v_{i,t},
$$
where $\beta \in [0,1]$ and $\gamma > 0$ are user-defined hyperparameters.
Theorem~\ref{thm:risk}
provides a bound for the MSE, assuming that  $\I(\bX_{1:t}^i;~X^j_{t} |
\bX^j_{1:t-1})$ is piecewise constant, and the estimate has i.i.d. noise with
bounded variance. We use the abbreviation $i_t = \I(\bX_{1:t}^i;~X^j_{t} |
\bX^j_{1:t-1})$, and similarly  $\hat{i_t} = \hat{\I}(\bX_{1:t}^i;~X^j_{t} |
\bX^j_{1:t-1})$ for convenience.

\begin{nthm}
  \label{thm:risk}
  Let $\hat{i}_t = i_t + \epsilon_t$, where $\epsilon_t$ is independent with
  mean 0 and variance $\sigma_t^2$, and $i_t$ is piecewise constant with $m$ transitions. Then the MSE
  of the ADI ensemble estimator is bounded by:
  \begin{multline}
    \EE\left[\sum_{t=1}^T (\overline{\textup{ADI}}(t) - i_t)^2 \right] \le
    \frac{m}{\gamma} \ln n_t \\ - \frac{1}{\gamma} \ln \beta^m (1 - \beta)^{T-m} + \frac{\gamma}{8}T + m\sigma_*^2\ln\left( \frac{T}{e} \right),
  \end{multline} 
where $\sigma_*^2 = \max_t \sigma_t^2$.
\end{nthm}
The proof of Theorem~\ref{thm:risk} is given in Appendix~\ref{sec:proof}.

\section{Spatial Interaction Estimation in a Scene}
\label{sec:agent_model}

We illustrate ADI by applying it to discover salient time-varying interactions among actors in
a scene. Here, the components $n=1, \ldots, N$ are actors moving around in
space. For each sampled frame $t$ and actor $i$, define the position vector $X^i_t =
[x^i_t, y^i_t]$ on the plane.  

\subsection{Dynamic Covariance Model}
\label{ssec:cov}

We propose a dynamic Gaussian model, following the model
in~\cite{chen2016dynamic}. Assume that the combined feature matrix is
distributed as:
\begin{equation}
  \label{eq:cov}
  \bX \sim \mathcal{N}(m_t, \Sigma_t),
\end{equation}
where $m_t$ is a mean vector and $\Sigma_t$ is a covariance matrix. We assume
that $m_t$ and $\Sigma_t$ are slowly varying, and further use a kernel
estimate of these quantities:
\begin{align}
  \label{eq:estimates}
  \hat m_t &= \frac{1}{\sum_{i=1}^T K_h(i - t)} \sum_{i=1}^T K_h(i - t) X_i. \\
  \hat \Sigma_t &= \frac{1}{\sum_{i=1}^T K_h(i - t)} \sum_{i=1}^T K_h(i - t) (X_i - \hat m_i)  (X_i - \hat m_i)^T,
\end{align}
where $K_h(t)$ is a kernel function.
The conditional mutual information is a function of the covariance matrices
under a Markovian Gaussian random process~\cite{}.
$$
\hat{\I}(\bX^i_{1:t};X^j_{t}| X^j_{t-1}, X_{t-1}^{[N]/\{i,j\}}) = \frac{1}{2}\log\frac{\left|  \hat{\Sigma}_{X^j_t |
    X^j_{t-1}, X_{t-1}^{[N]/\{i,j\}}} \right|}{\left|  \hat{\Sigma}_{X^j_t |
    X^j_{t-1},  X^i_{t-1} , X_{t-1}^{[N]/\{i,j\}}}\right|}.
$$

\section{Application to Stanford Drone Dataset}
\label{sec:exp}

In this section, the proposed ensemble ADI estimator is applied to the Stanford Drone
Dataset~\cite{robicquet2016learning}, which  is a collection of 60 annotated videos
across 8 scenes shot on the Stanford campus. These annotations allow for tracking the
movement of pedestrians, cars, bicyclists and other moving actors in the scene.
These estimated locations of actors are smoothed by a moving mean estimator in
order to reduce artifacts introduced by
the discretization of the annotations. These smoothed locations for each actor
in the scene are then used to calculate the ADI.

For the analysis, an rbf kernel was used in (\ref{eq:estimates}) with parameter $h=5$, and the ADI
ensemble parameters were set to $\tau=10,\beta=0.01,\gamma=1$, and $G =
\{\text{exp}(0.1), \text{exp}(0.2), \text{unif}\}$.  After calculating ADI, only
interactions where the actors were within a certain distance (in pixels) from
each other were considered - in this case, 100.

\subsection{Interaction Example between Pedestrians}

\begin{figure}[h]
  \centering
  \includegraphics[width=3.38in]{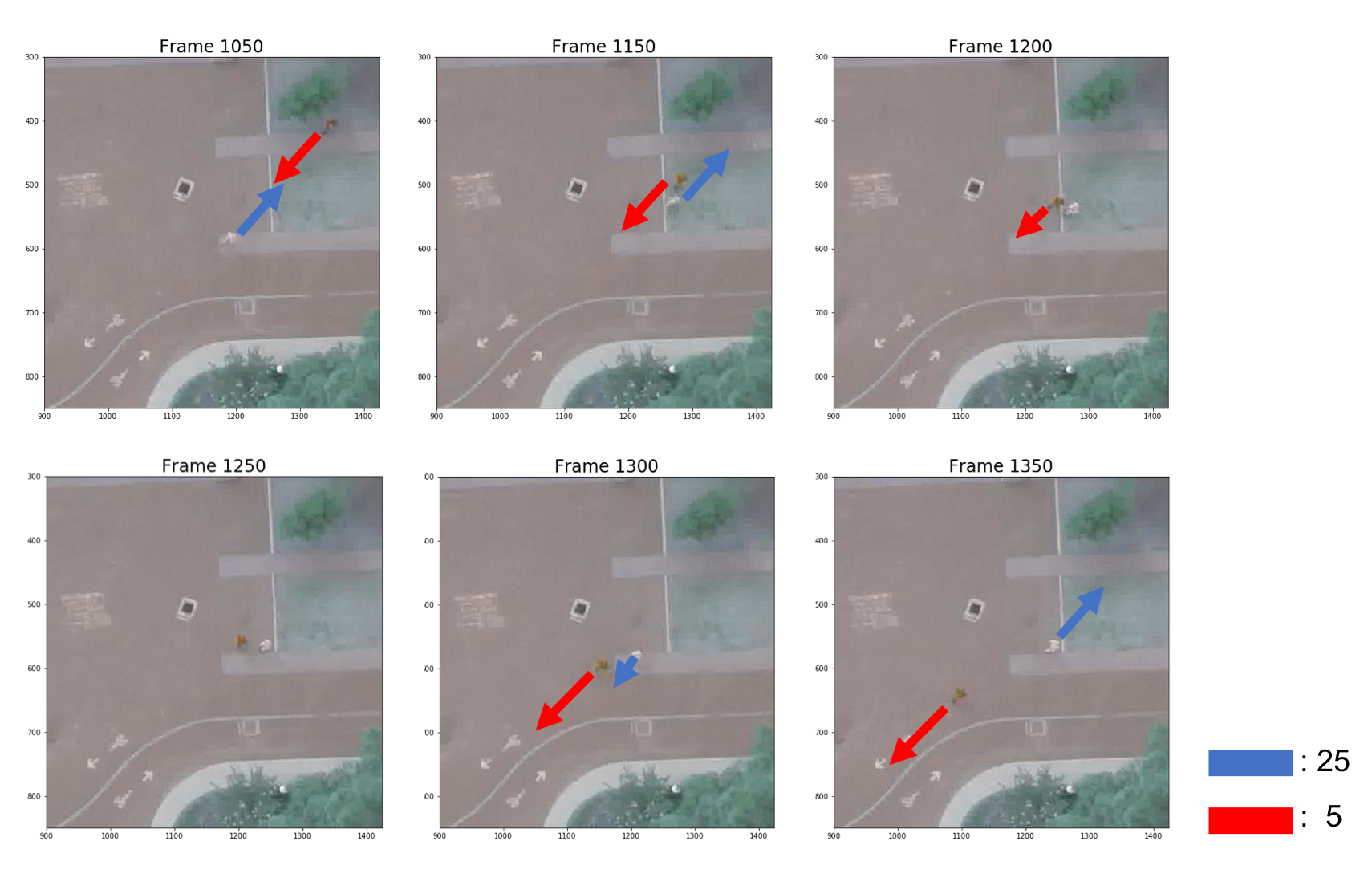}
  \includegraphics[width=2.5in]{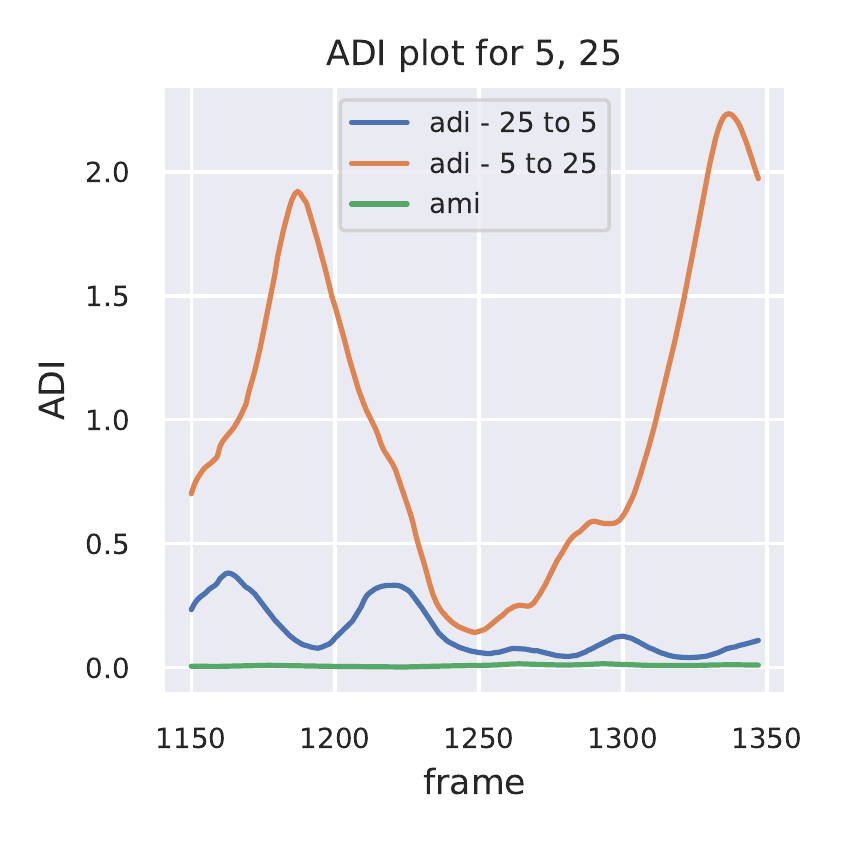}
  \caption{Stanford video dataset example. Here, we capture an interaction of
    two people meeting in the scene. The shown video frames and corresponding
    ADI are demonstrating them coming towards each other, interacting briefly,
    as actor 25 even walks in the other direction to continue the conversation,
    and then resuming their original path.  The title on the
  plots pair of labeled actors in ``video0'' of the bookstore scene,
  and the line labeled $i$ to $j$ represents $\ADI^{i \rightarrow j}$.}
  \label{fig:stanford}
\end{figure}

Fig.~\ref{fig:stanford} shows one example of ADI and the corresponding
interaction between two pedestrians. The pedestrians labeled 5 and 25 stop to
chat briefly, with 25 actually reversing course for a small time to continue the
conversation at frame 1280 to 1300 to continue the conversation. The estimated
ADI is able to identify this interaction, and to identify that there is more influence
from 5 to 25 than vice versa over this small window. This is compared with an
adaptive version of mutual information:
$$
\textrm{AMI}(\bX_{1:T}^i, \bX_{1:T}^j) = \sum_{t=1}^T
g^*(t,T)\hat{\I}(X_{t}^i;X^j_{t} | \bX^i_{1:t-1}, \bX^j_{1:t-1}),
$$

where the ensemble method outlined for ADI is applied to the estimated summand
$\hat{\I}(X_{t}^i;X^j_{t} | \bX^i_{1:t-1}, \bX^j_{1:t-1})$.
\subsection{Visualization of Interactions based on ADI}

\begin{figure}[h]
  \centering
  \includegraphics[width=3.38in]{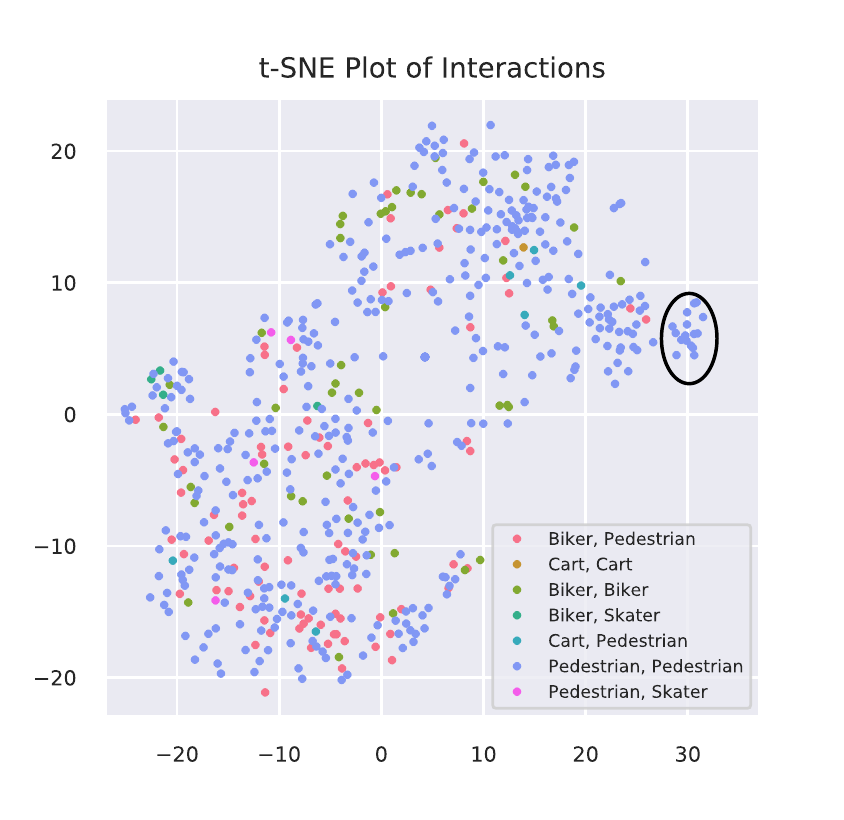}
  \caption{t-SNE plot of interactions based on ADI. The highlighted cluster of
    pedestrian interactions is characterized by low levels of interaction over a
  long period of time combined with spikes of activity.}
  \label{fig:tsne}
\end{figure}

We can use ADI as a tool to cluster and visualize many interactions in the dataset. First, the ADI
for all interactions between actors in the bookstore scene from the Stanford
Drone dataset across 5 different
videos are collected, totaling $m=539$ interactions. Using symmetrized ADI,
$\ADI^{i,j} = \ADI^{i\rightarrow j} +
\ADI^{j \rightarrow i}$, the maximal cross correlation between each interaction
is found, and this correlation is used as an affinity measure $a_{k,l}$, with the
corresponding affinity matrix $A = [a_{k,l}]_{k,l=1,\ldots m}$. Note that $a_{k,l} = a_{l,k}$, and so $A$ is
symmetric. $A$ can then be used to apply a number of visualization and clustering
techniques. Here, we use t-SNE dimension reduction and visualization
method~\cite{maaten2008visualizing}, by
transforming $A$ to a distance matrix $D = [d_{i,j}]_{i,j=1,\ldots m}$,
where $d_{i,j} = \sqrt{2(1 - a_{i,j})}$ and applying the method to this
matrix. Fig.~\ref{fig:tsne} shows the results. The colors correspond to
different types of interactions, such as between pedestrians, or between a
pedestrian and a bike, etc.

\begin{figure}[h]
  \centering
  \includegraphics[width=3.38in]{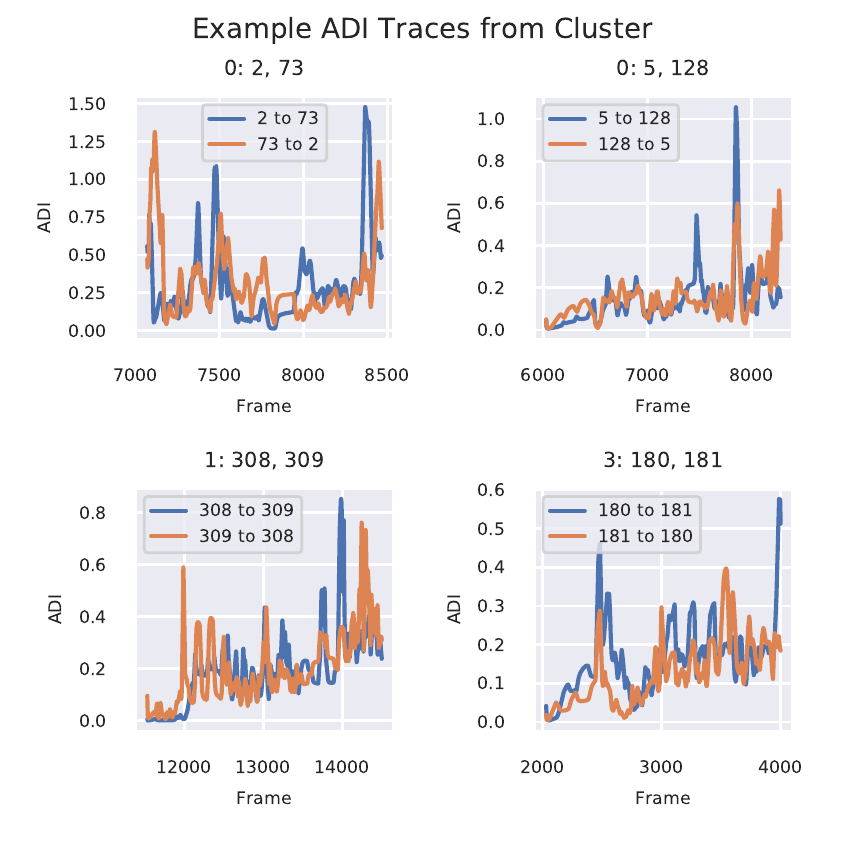}
  \caption{Representative ADI traces from highlighted cluster. The majority of these
  interactions  are pedestrians that are moving slowly together or standing
  still in close proximity, with abrupt direction and velocity changes. The titles on the
  plots represent the origin video and pair of labeled actors in the dataset,
  and the line labeled $i$ to $j$ represents $\ADI^{i \rightarrow j}$.}
  \label{fig:clus_example}
\end{figure}

The visualization shows small clusterings of interactions. An example is circled
in black, with representative traces shown in Fig.~\ref{fig:clus_example}.
More generally, we see that the pedestrian-biker interactions mostly cluster in
the bottom-left portion of the plot, while the biker-biker and
pedestrian-pedestrian interactions are less cohesive as a group, implying
heterogeneity among these types of interactions. The small 
highlighted cluster of pedestrian interactions, for example, are characterized
by long periods of low ADI combined with abrupt spikes. These are observed to
correlate to pedestrians walking slowly in the same direction or standing still
along with occasional changes in velocity or direction. 

\subsection{Relationship between ADI and Velocity}

In this section we study the relationship between the velocity profile and ADI
profile of particular types of interactions. For each interaction and each actor
$i$ the instantaneous velocity vector $\mathbf{v}^i_t = [v^i_{t,x}, v^i_{t,y}]$
is calculated, along with the
corresponding instantaneous magnitude $v^i_t = \left\lVert \mathbf{v}^i_t
\right\rVert$. Further, the instantaneous velocity angle between two actors $i$
and $j$ is calculated:
$$
\theta^{i,j}_t = \arccos\left( \frac{\mathbf{v}^i_t \cdot \mathbf{v}^j_t}{v^i_t v^j_t} \right).
$$

\begin{figure}[h]
  \centering
  \includegraphics[width=3.38in]{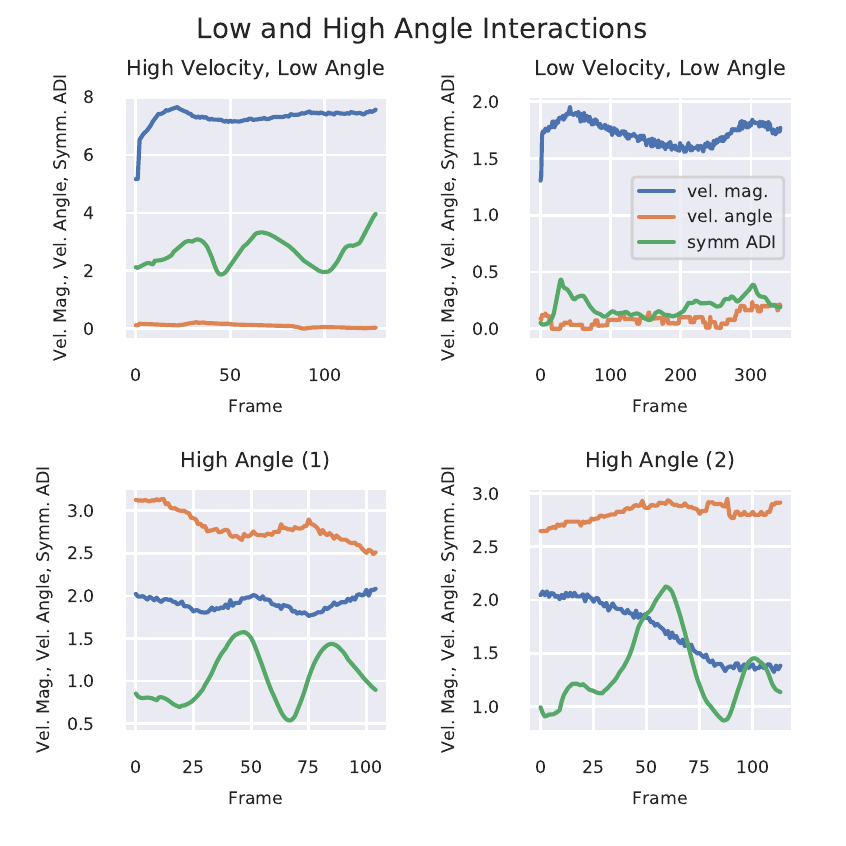}
  \caption{Representative profiles of low and high velocity angle interactions.
    The top row shows two low-angle interactions, one with high total velocity.
    The high total velocity interaction has a relatively constant symmetrized ADI profile,
    while the low total velocity interaction has an ADI profile close to 0. The
    high angle interactions have more variable ADI profiles relative to their
    magnitude, and tend to be sensitive to changes in total velocity.}
  \label{fig:low_high}
\end{figure}

Using the relative velocity angle, we can look for two specific types of
interactions, and how their ADI profiles differ; those with high angle, so that the two actors are
approaching from opposite directions, and low angle, where the two
actors are moving in the same direction. Fig.~\ref{fig:low_high} shows four
representative interactions, two with low velocity angles and two with high
velocity angles.

In general, interactions with high total velocity, defined as $v^i_t
+ v^j_t$, and low velocity angle see a stable and non-zero symmetrized ADI. In
the low total velocity setting, the ADI is normally much smaller than
its high velocity counterpart. Two examples of low-angle interactions are shown
in the top row of Fig.~\ref{fig:low_high}. In the high angle case, ADI is less constant, and
in many cases responds more to changes in total velocity, as shown in the bottom
row of Fig.~\ref{fig:low_high}.

\subsection{Average ADI between Different Types of Actors}
\begin{figure}[h]
  \centering
  \includegraphics[width=3.38in]{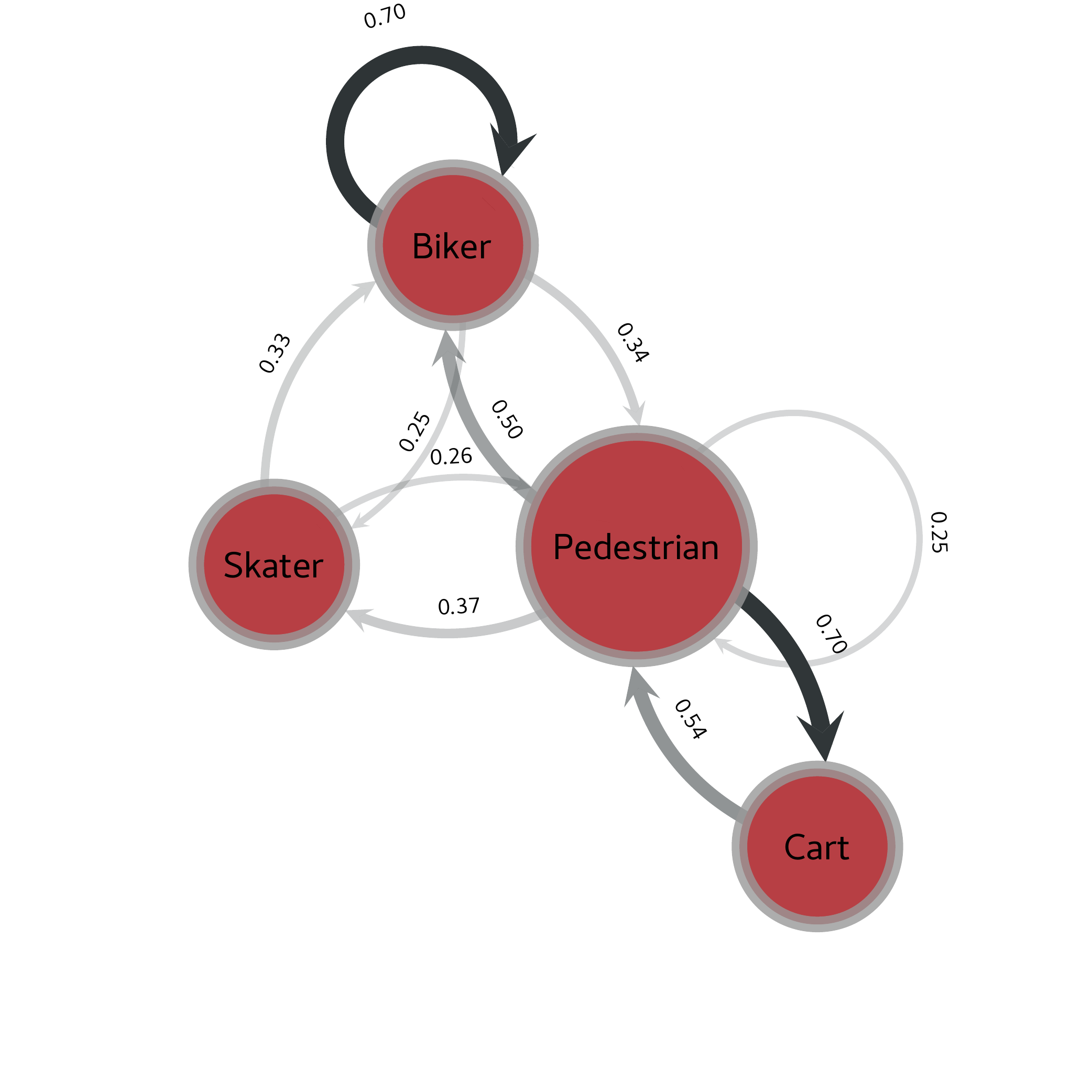}
  \caption{Average ADI between types of actors in the bookstore scene for the
    Stanford drone dataset. Bikers have the largest levels of interaction, while
    skaters have the least.}
  \label{fig:bookstore_heatmap}
\end{figure}

Fig.~\ref{fig:bookstore_heatmap} shows a graph of the average ADI between types of actors
in the bookstore scene from the Stanford drone dataset across 5 different
videos.

Skaters tend to have the lowest average ADI with other groups, followed by
pedestrians, with bikers and carts having the largest interaction magnitudes. Interestingly,
pedestrians influence bikers and carts more than the two groups influence pedestrians on average,
possibly signifying that bikers and carts are more cautious and thus are more affected by pedestrians in the vicinity. As seen in
Fig.~\ref{fig:low_high}, the velocity magnitudes in interactions can play a
role, specifically that the magnitudes of velocity and ADI are positively correlated.
With bikers being among the fastest moving actors in this graph, it makes sense that
they have some of the largest interaction magnitudes. %

\section{Conclusion}
\label{sec:concl}
In this paper, we introduced an ADI estimator that utilizes an ensemble
technique in order to make ADI more robust to user-specified parameters. The
estimator is applicable to  
real-world scenarios where directed information evolves as a function of time.
We illustrated the power of the ensemble ADI estimator to detect
latent interactions in a video using the
Stanford drone dataset. In the future, ADI can be used as a data summarization and
exploration tool or as a component in a larger system.

\appendix
\section{Proof of Theorem 2.1}
\label{sec:proof}
To aid in the proof, we prove two propositions and restate Theorem 2
in~\cite{shalizi2011adapting} as Lemma~\ref{lem:regret}. Since we assume that $i_t$ is piecewise constant
with $m$ changes, we can define $[t_1, t_2, \ldots, t_m]$ as the (unknown)
transition points, where $t_1=1$. We further define the ``oracle mean
estimator'' $u^*(t)$:
\begin{equation*}
 u^*(t) = \sum_{k=1}^m \sum_{t=1}^T \frac{1}{t - t_k + 1} \mathbf{1}(t_k \le t < t_{k+1})i_t, 
\end{equation*}
where $t_{m+1} = T + 1$. The proof is based on the following result found
in~\cite{shalizi2011adapting}, restated in terms of ADI:
\begin{nlem}
  \label{lem:regret}
  The tracking regret of the ensemble ADI estimator in comparison with $u^*(t)$, defined as:
  \begin{equation*}
    R(u^*(T)) = \sum_{t=1}^T (\overline{\textup{ADI}}(t) - i_t )^2 - \sum_{t=1}^T (u^*(t) - i_t )^2,
  \end{equation*}
  is at most
  \begin{equation}
     R(u^*(T)) \le \frac{m}{\gamma}\ln n_t - \frac{1}{\gamma}\ln \beta^m (1 - \beta)^{T-m} + \frac{\gamma}{8}T.
   \end{equation}
\end{nlem}

\begin{nprop}
  \label{prop:oracle_regret}
    \begin{equation}
      \EE \left[ (u^*(t) - i_t)^2 \right] \le \sigma_t^2 + \frac{1}{t - t_k + 1}\sigma_*^2.
    \end{equation}
  \end{nprop}

  \begin{proof}
    \begin{align*}
      \EE \left[  (u^*(t) - i_t)^2 \right] &= \EE \left[ \left( \frac{1}{t - t_k + 1}\sum_{i=t_k}^{t} (\theta_i + \epsilon_i) - (\theta_t + \epsilon_t) \right)^2 \right] \\ 
      &= \EE \left[ \left( \frac{1}{t - t_k + 1}\sum_{i=t_k}^{t-1} \epsilon_i -  \epsilon_t \right)^2 \right] \\
       &= \frac{1}{(t - t_k + 1)^2} \left( \sum_{i=t_k}^{t-1} \epsilon_i + (t - t_k)^2 \epsilon_t  \right) \\
      &\le   \frac{(t - t_k)}{(t - t_k + 1)^2}\sigma^2_{*}  + \frac{(t - t_k)^2}{(t - t_k + 1)^2}\sigma_t^2 \\
      &\le  \frac{\sigma_*^2}{t - t_k + 1} + \sigma_t^2.
      \end{align*}
\end{proof}

\begin{nprop}
  \label{prop:ADI_regret}
  \begin{equation}
    \EE\left[(\overline{\textup{ADI}}(t) - i_t)^2\right] \ge  \EE\left[ (\overline{\textup{ADI}}(t) - \theta_t)^2 \right] + \sigma_t^2.
  \end{equation}
\end{nprop}

\begin{proof}
  We first decompose the left side using the definition of $i_t$:
  $$
  (\overline{\ADI(t)} - i_t)^2 = (\overline{\ADI}(t) - \theta_t)^2 + 2\epsilon_t(\overline{\ADI}(t) - \theta_t) +
  \epsilon_t^2.
  $$
  The result follows from taking the expectation of both sides, along with the
  following observation:
  \begin{align}
    \EE \left[ 2\epsilon_t(\overline{\ADI}(t) - \theta_t)  \right]
    &= \EE \left[ 2\epsilon_t \sum_{j=1}^{n_t} w_{j, t-1} \sum_{i=1}^t g_j(i,T; t_0) i_i  \right] \\
    &= \EE \left[ 2\sum_{j=1}^{n_t} w_{j,t-1} g_j(i,T; t_0) i_t \epsilon_t \right]\\
    &= 2\sigma_t^2\sum_{j=1}^{n_t} w_{j,t-1} g_j(i,T; t_0) \ge 0,
  \end{align}
  where the last inequality is due to the fact that $w_{j,t-1}$ and
  $g_j(i,T;t_0)$ are non-negative, $\forall i, j, t$.
  \end{proof}

Using the definition of $R(u^*(T))$ and Props.~\ref{prop:ADI_regret},
\ref{prop:oracle_regret}, we obtain:
\begin{multline*}
 \sum_{t=1}^T \left(  \EE\left[ (\overline{\textup{ADI}}(t) - \theta_t)^2 
   \right] + \sigma_t^2 \right) - \\
  \sum_{t=1}^T \left(  \sigma_t^2 + \sum_{k=1}^m\frac{1}{t - t_k + 1} \mathbf{1}(t_{k} \le t < t_{k+1})\sigma_*^2 \right) \le R(u^*(T)).
\end{multline*}
Finally, note the following inequality:
\begin{equation*}
  \sum_{k=1}^m \sum_{t=1}^T \frac{1}{t - t_k + 1} \le m\ln\left( \frac{T}{e} \right).
\end{equation*}
Combining this with Lemma~\ref{lem:regret}, and rearranging terms, achieves the
desired bound. 
\printbibliography

\end{document}